\documentclass[preprint,review,12pt]{elsarticle}

\usepackage{amssymb}
\usepackage{amsmath}

\journal{}

\begin{document}

\begin{frontmatter}


\title{Chemical Short-Range Order Regulates Hydrogen Energetics and Hydrogen-Dislocation Interactions in CoNiV}

\author[label1]{Beihan Chen}
\author[label1]{Dalia Sayed Ahmed}
\author[label2]{Yang Yang}
\author[label1]{Miaomiao Jin\corref{cor1}}
\ead{mmjin@psu.edu} 

\cortext[cor1]{Corresponding author}

\affiliation[label1]{organization={Department of Nuclear Engineering, Pennsylvania State University},
             city={University Park},
             postcode={16802},
             state={PA},
             country={USA}}

\affiliation[label2]{organization={Department of Engineering Science and Mechanics, Pennsylvania State University},
             city={University Park},
             postcode={16802},
             state={PA},
             country={USA}}


\begin{abstract}
Chemical short-range order (CSRO) has emerged as a critical structural feature in concentrated alloys, yet its coupling with hydrogen remains an active discussion. Here, we develop a machine-learning interatomic potential for the Co–Ni-V–H system and investigate how CSRO regulates hydrogen energetics and dislocation behavior in CoNiV, an alloy with reported strong resistance to hydrogen embrittlement. We identify strong V-centered ordering that suppresses V–V clustering and significantly reshapes the hydrogen solution landscape. Compared to a chemically random alloy, the ordered state exhibits higher average hydrogen solution energies and a reduced population of strongly binding sites, indicating lower bulk hydrogen uptake. At partial dislocations, hydrogen preferentially segregates to tensile core regions, acting as a shallow, reversible trap with a much weaker effect compared to chemical trapping states. These results demonstrate that local chemical order strongly regulates hydrogen–dislocation coupling and provide an atomistic understanding for tuning hydrogen-assisted deformation in concentrated CoNiV alloys.
\end{abstract}


\end{frontmatter}



\section{Introduction}
Multi-principal element alloys (MPEAs), including medium- and high-entropy alloys, have attracted significant interest due to their ability to form single-phase solid solutions with exceptional mechanical properties that often surpass those of conventional alloys \cite{ borg2020expanded,senkov2015accelerated,tsai2014high}. Among these, equiatomic or near-equiatomic ternary alloys such as VCoNi represent a particularly attractive model system: despite combining elements with distinct chemical and physical characteristics, they can (meta)stabilize an FCC solid solution with high strength and damage tolerance \cite{sohn2020high,sohn2019ultrastrong,tian2021effects}. Although MPEAs are often treated as chemically random solid solutions, a growing body of experimental and computational work has demonstrated the presence of chemical short-range order (CSRO) in many MPEA systems \cite{singh2015atomic}. CSRO introduces locally heterogeneous chemical environments that may influence defect energetics, solute transport, and mechanical response \cite{jian2020effects,yin2020yield}. In the Co–Ni-V system, prior studies have reported a tendency for preferential Ni–V and Co–V nearest-neighbor pairing and avoidance of V–V pairs, indicating pronounced V-centered CSRO \cite{kostiuchenko2020short,chen2021direct}.

Separately, recent experiments have shown that CoNiV alloys exhibit high resistance to hydrogen embrittlement under ambient conditions \cite{luo2020strong}. Hydrogen embrittlement remains a persistent challenge for structural alloys exposed to hydrogen-containing environments, arising from hydrogen interactions with lattice defects that degrade ductility and fracture resistance \cite{laureys2018role,pressouyre1980trap}. Extensive studies in metals and alloys have established that hydrogen–defect interactions govern embrittlement behavior, motivating mechanisms such as hydrogen-enhanced decohesion and hydrogen-enhanced localized plasticity \cite{li2022hydrogen,yu2024hydrogen}. Hydrogen trapping strongly or weakly at sites like vacancies, dislocations, interfaces, and cavities plays a central role in controlling hydrogen transport by affecting the concentration of diffusible hydrogen \cite{pressouyre1980trap,laureys2018role,chen2020observation,sato2023quantitative}; dislocations are crucial hydrogen trapping sites due to their direct involvement in plastic deformation and interactions with other lattice defects. In addition, hydrogen effects are highly sensitive to local chemical environments, leading to material-dependent and sometimes conflicting observations regarding hydrogen diffusivity and its impact on dislocation activity and mechanical strength \cite{uhlemann1998diffusivity,husby2018effect,hanson2018crystallographic,bond1988effects,ichii2018comparative,harris2018elucidating,ogawa2020hydrogen,robertson1986hvem}.

While hydrogen behavior has been extensively studied in pure metals and conventional alloys, its interaction with local chemical order in MPEAs is far less understood. In particular, how CSRO influences hydrogen solution energetics and hydrogen–defect interactions in Co–Ni-V alloys is still an open question. Luo et al. attributed the observed hydrogen embrittlement resistance to a combination of factors, including a possible slight stacking fault energy reduction from hydrogen, thereby promoting nanotwin formation and enhanced local strain hardening, as well as low hydrogen diffusivity and reduced hydrogen uptake associated with a dense oxide film \cite{luo2020strong}. However, the coexistence of hydrogen resistance and pronounced CSRO raises fundamental questions regarding how CSRO affects hydrogen solution thermodynamics and hydrogen trapping at dislocations, with potential implications for macroscopic mechanical behavior. Given the  V-centered CSRO identified in CoNiV alloys \cite{chen2020observation,sato2023quantitative}, prior studies of vanadium-containing systems (with H) provide important context. Vanadium exhibits strong hydrogen affinity and high hydrogen solubility in its pure form \cite{owen1972relation}, while vanadium-rich environments in alloys can act as deep hydrogen traps \cite{li2018effects}, such as at the surfaces of vanadium carbide particles (V$_4$C$_3$) \cite{takahashi2012direct,takahashi2018origin} and within vanadium-enriched carbides in steels \cite{cheng2018hydrogen}. It remains unclear how established hydrogen–vanadium interactions translate to V-centered CSRO in CoNiV alloys.

First-principles methods such as density functional theory (DFT) provide accurate descriptions of hydrogen energetics and bonding \cite{nazarov2014ab,hickel2014ab} but are computationally prohibitive for the large systems required to study CSRO and dislocation cores. Conventional empirical potentials, while computationally efficient, often lack the transferability needed to describe multi-component alloys. Machine-learning interatomic potentials (MLIPs) offer a powerful alternative by reproducing DFT-level accuracy while enabling large-scale simulations of complex materials systems \cite{behler2007generalized,kwon2023accurate,mishin2021machine}. In this work, we develop and validate an MLIP for the Co-Ni-V-H system. Using this potential, we investigate (i) the nature of CSRO in equiatomic VCoNi with/without hydrogen, (ii) hydrogen solution energetics in systems with/without CSRO, and (iii) hydrogen interactions with partial dislocations. Those results provide atomic-scale insight into hydrogen behavior in VCoNi alloys.

\section{Method}
\subsection{Development of the MLIP}
An MLIP for the Co–Ni-V–H system was developed using a DFT database constructed to sample a broad range of chemical and local atomic environments relevant to bulk alloys, CSRO, and hydrogen-containing configurations. 
\begin{itemize}
    \item \textbf{Multicomponent alloy configurations:}  FCC supercells ($3 \times 3 \times 3$, 108 atoms) with random chemical distributions were generated under the constraints $\text{Ni} > 33\%$, $\text{Co} > 10\%$, and $\text{V} > 10\%$ to target the stability range of the FCC phase. Additional configurations exhibiting CSRO were included, along with structures strained near the equilibrium volume to better capture the energetics of local lattice distortion.
    \item \textbf{Hydrogen-bearing configurations:} Hydrogen atoms (2-5 atoms) were inserted into both tetrahedral and octahedral interstitial sites within alloy configurations. This sampling accounted for the diverse local chemical environments in the alloy.
    \item \textbf{Finite-temperature configurations:} Additional snapshots were extracted from finite-temperature molecular dynamics (MD) simulations at elevated temperatures to capture thermally activated lattice distortions and transient hydrogen residence not accessible through static sampling.
\end{itemize}

All DFT calculations were performed using the Vienna Ab initio Simulation Package (VASP 6.4.3) with the projector augmented-wave (PAW) method and the Perdew–Burke–Ernzerhof (PBE) generalized gradient approximation. A plane-wave cutoff energy of 520 eV was used, with PAW potentials Ni\_pv, Co\_sv, V\_sv, and H. Brillouin-zone sampling employed a $2 \times 2 \times 2$ Monkhorst–Pack $k$-point mesh for the supercells. Electronic convergence was set to $10^{-3}$ eV for total energies, corresponding to $\sim$0.01 meV/atom accuracy, and atomic forces were collected for all configurations.

The MLIP was trained using the moment tensor potential (MTP) framework \cite{Shapeev2016}, in which the total energy is expressed as a sum of environment-dependent atomic energies constructed from local moment tensor descriptors within a finite cutoff radius (5.0 {\AA}). Model parameters were optimized by minimizing the errors in energies and atomic forces relative to the DFT reference data. Initial potentials were used in finite-temperature MD simulations to identify extrapolative configurations, which were subsequently added to the DFT training set for further potential training. This iterative procedure was repeated until no significant extrapolation was detected. A similar workflow has been successfully applied in our previous work on the uranium nitride system \cite{Chen2025UNpot}. The final training dataset comprised 6,394 configurations, and a schematic of the training workflow is provided in Supplementary Materials (SM) Figure S1.

The accuracy and transferability of the MLIP were assessed using a series of standard validation tests. These included evaluation of total energy and atomic force prediction errors on a validation set not used during training, as well as calculations of equilibrium lattice constants and elastic constants for pure FCC Ni and equiatomic VCoNi structures, benchmarked against previous DFT reference data and experimental data. Chemical mixing behavior was examined by testing whether the potential reproduces expected energetic preferences among near-neighbor atomic pairs as reported in previous studies. Hydrogen-related validation focused on site-dependent solution energetics in FCC environments and hydrogen interactions with point defects, using hydrogen–vacancy binding in pure Ni as a test case with well-established DFT reference data. These validation steps establish the suitability of the potential for simulations of chemical ordering, hydrogen solution, and hydrogen–defect interactions in CoNiV alloys.

\subsection{MD Simulation Procedures}
MD simulations were carried out using the LAMMPS package \cite{plimpton1995fast} with the validated MLIP. CSRO in CoNiV was investigated using Monte Carlo (MC) MD simulations. Simulations were carried out on an $8 \times 8 \times 8$ FCC supercell containing 2,048 equimolar metal atoms over temperatures ranging from 300 to 1,000 K in 100 K increments. Each simulation was initialized from a chemically random configuration, and atomic swaps between V, Co, and Ni were accepted or rejected using the Metropolis algorithm at each temperature. For each temperature, $10^6$ attempted swaps were performed, and simulations were repeated six times to ensure statistical reliability. Pair correlation statistics were used to compute first-nearest-neighbor Warren–Cowley short-range order parameters $\alpha_{ij}(1)$ for all element pairs \cite{cowley1950approximate}. The influence of hydrogen on CSRO was examined by introducing 1 at.\% hydrogen and repeating the same MCMD procedure.

Hydrogen solution energies, $E_{\text{sol}}$, were evaluated in the metal supercells where a single hydrogen atom was introduced into the octahedral interstitial site, followed by full atomic relaxation. The hydrogen solution energy was defined as,
\begin{equation}
    E_{\text{sol}} = E(\text{NiCoV+H}) - E(\text{NiCoV}) - \frac{1}{2}E(\text{H}_2)
\end{equation}
where $E(\mathrm{VCoNi}+H)$ and $E(\mathrm{VCoNi})$ denote the total energies of the supercell with and without hydrogen, respectively, and $E(\mathrm{H}_2)=-6.77$ eV is the DFT-computed energy of an isolated H$_2$ molecule, with inter-atomic distance of 0.75 {\AA}, consistent with prior calculations \cite{connetable2015first,connetable2014segregation}. All energetic results are reported without zero-point energy correction. 

To investigate hydrogen interactions with dislocations, an edge dislocation model was constructed in the FCC alloy. A $a/2\langle110\rangle{111}$ edge dislocation dipole was introduced into an orthorhombic supercell oriented along $[110]\times[\bar{1}12]\times[1\bar{1}1]$, containing 28,392 atoms, with periodic boundary conditions applied in all directions. Atomic positions were first relaxed by energy minimization using the ML potential, followed by low-temperature equilibration at 10 K under isothermal-isobaric conditions (zero external pressure), with the simulation cell dimension along the dislocation line direction held fixed, to obtain a stable dissociated dislocation core structure. Starting from this relaxed dislocation configuration, hydrogen solution energetics were evaluated as a function of distance from the dislocation core by placing individual hydrogen atoms at octahedral interstitial sites and minimizing energy to compute local solution energies.

\section{Results and Discussion}
\subsection{Potential Validation}
\begin{figure}
    \centering
    \includegraphics[width=0.85\linewidth]{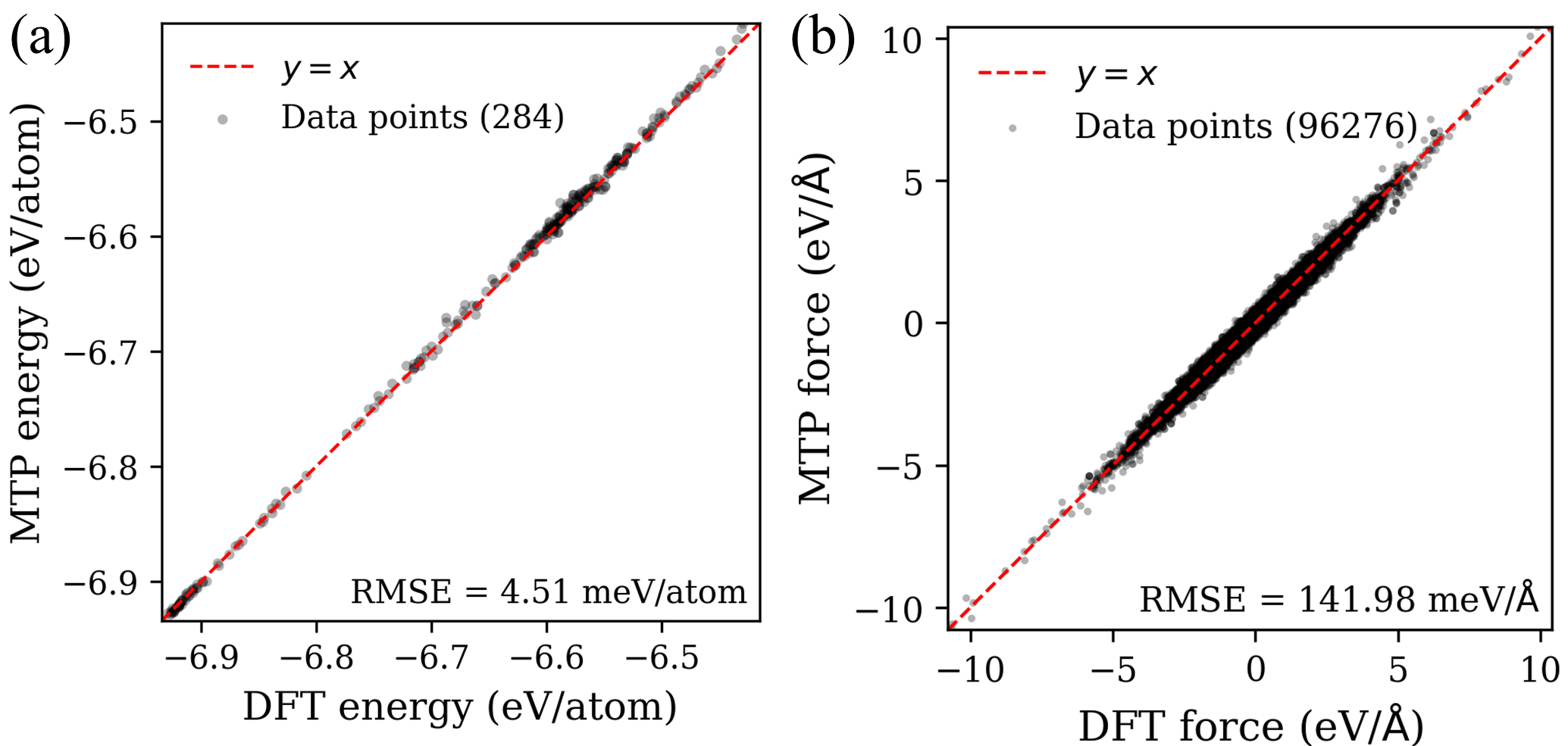}
    \caption{Validation of MLIP-predicted energies and forces against DFT reference data for the validation set.}
    \label{fig:rmse}
\end{figure}

The accuracy of the trained MLIP was first evaluated through direct comparison of predicted energies and atomic forces against DFT values for a validation dataset not used during training. Figure \ref{fig:rmse} depicts the validation performance for a validation set comprising 284 configurations. The root-mean-square error (RMSE) in energy is 4.52 meV/atom. The force prediction accuracy, evaluated over 96,276 atomic force components, yields an RMSE of 141.98 meV/{\AA}. These energy and force error levels are comparable to, or better than, those reported in prior MLIP developments for chemically complex systems \cite{schmidt2019recent}, e.g., around 3 meV/atom and 0.05-0.2 eV/{\AA} for Mo-Nb-Ta-V-W alloys by Byggm{\"a}star et al. \cite{byggmastar2021modeling}. Hence, the present MLIP captures the underlying DFT potential energy surface with sufficient fidelity for atomistic simulations. Based on our testing (SM Table S1), increasing model complexity beyond the chosen potential form did not result in statistically meaningful improvements in accuracy, and the final model was selected to balance predictive performance and computational efficiency. The resulting potential is therefore well-suited for large-scale MD simulations and available to the readers.

Beyond energy and force metrics, the MLIP was further validated against fundamental structural and mechanical properties. Table \ref{tab:elastic_constants} compares lattice constants and elastic constants of pure FCC Ni and equiatomic CoNiV obtained using the MLIP with available experimental and DFT reference data. Lattice parameters and elastic constants were calculated following standard finite-strain procedures, as described in SM Section S3. For pure Ni, the MLIP reproduces the experimental lattice constant and elastic constants with good agreement, providing a baseline validation for elemental bonding and mechanical response. For equiatomic CoNiV, the MLIP predictions are consistent with reported experimental measurements and prior DFT calculations, demonstrating that the potential well captures both equilibrium structure and elastic response in the chemically complex alloy.

\begin{table}[htbp]
\centering
\small
\caption{Lattice parameters and elastic constants (unit: GPa) of Ni and equiatomic CoNiV alloys. Random solid-solution configurations were used for alloy calculations.}
\begin{tabular*}{\textwidth}{@{\extracolsep{\fill}}llllllll}
\hline
& $a$ (\AA) & $C_{11}$  & $C_{12}$  & $C_{44}$ & $B$ & $E$  & $G$  \\
\hline
Exp. Ni \cite{Alers1960} & 3.52 & 261.2 & 150.8 & 131.7 & 187.6 & 239.4 & 92.9 \\
DFT Ni (this work)  & 3.52 & 282.2 & 150.6 & 138.6 & 194.5 & 262.1 & 102.8 \\
MTP Ni (this work)   & 3.51 & 251.4 & 160.1 & 141.8 & 190.5 & 233.5 & 90.2 \\
\hline
Exp. CoNiV \cite{sohn2019ultrastrong} & 3.60 & - & - & - &  & 192 & 72 \\
DFT CoNiV \cite{zhou2021mechanism} & 3.581 & 251 & 173 & 109 & - & - & - \\
DFT CoNiV (this work) & 3.59 & 289 & 165 & 131 & 206 & 251 & 97 \\
MTP CoNiV (this work)  & 3.58 & 266 & 189 & 128 & 214 & 211 & 79 \\
\hline
\end{tabular*}
\label{tab:elastic_constants}
\end{table}

To validate H-metal interactions described by the MLIP, we compare our calculations of hydrogen energetics in Ni against established DFT and experimental data from the literature \cite{nazarov2014ab}. Two representative test cases are examined: hydrogen solution in the pristine Ni lattice and hydrogen binding to a Ni monovacancy. The hydrogen solution energy is calculated to be 0.18 eV for the octahedral interstitial site, which is the energetically preferred site in FCC Ni. This value is generally comparable to the previous DFT result of 0.11 eV \cite{metsue2016hydrogen}, 0.05 eV \cite{nazarov2014ab} and 0.09 eV \cite{connetable2014segregation} for hydrogen at the same site. The differences likely arise from variations in first-principle settings, such as exchange–correlation functional, supercell size, and k-point sampling. For hydrogen in a vacancy, we note that a single H atom does not favor the vacancy center, rather it tends to adsorb on the inner surface of the vacancy. When hydrogen is placed at the vacancy center, we obtain a binding energy of -0.61 eV, indicating a strongly unfavorable configuration and closely matching the reported DFT value of -0.65 eV \cite{nazarov2014ab} and -0.594 eV \cite{connetable2014segregation}. However, energy minimization of the off-center hydrogen-vacancy system yields a binding energy of 0.29 eV. This positive binding energy indicates a significant affinity of hydrogen for the vacancy and is in excellent agreement with published values (0.22 eV  \cite{nazarov2014ab} and 0.26 eV \cite{connetable2014segregation} for H trapped at a Ni monovacancy). Notably, this falls close to the experimental binding energy measurements (e.g., 0.24 eV, 0.27 eV, and 0.28 eV \cite{nazarov2014ab} and references within) associated with hydrogen in monovacancies in Ni. 
 
\subsection{CSRO and Hydrogen in Bulk}
\begin{figure}[htb]
\centering
\includegraphics[width=1.0\textwidth]{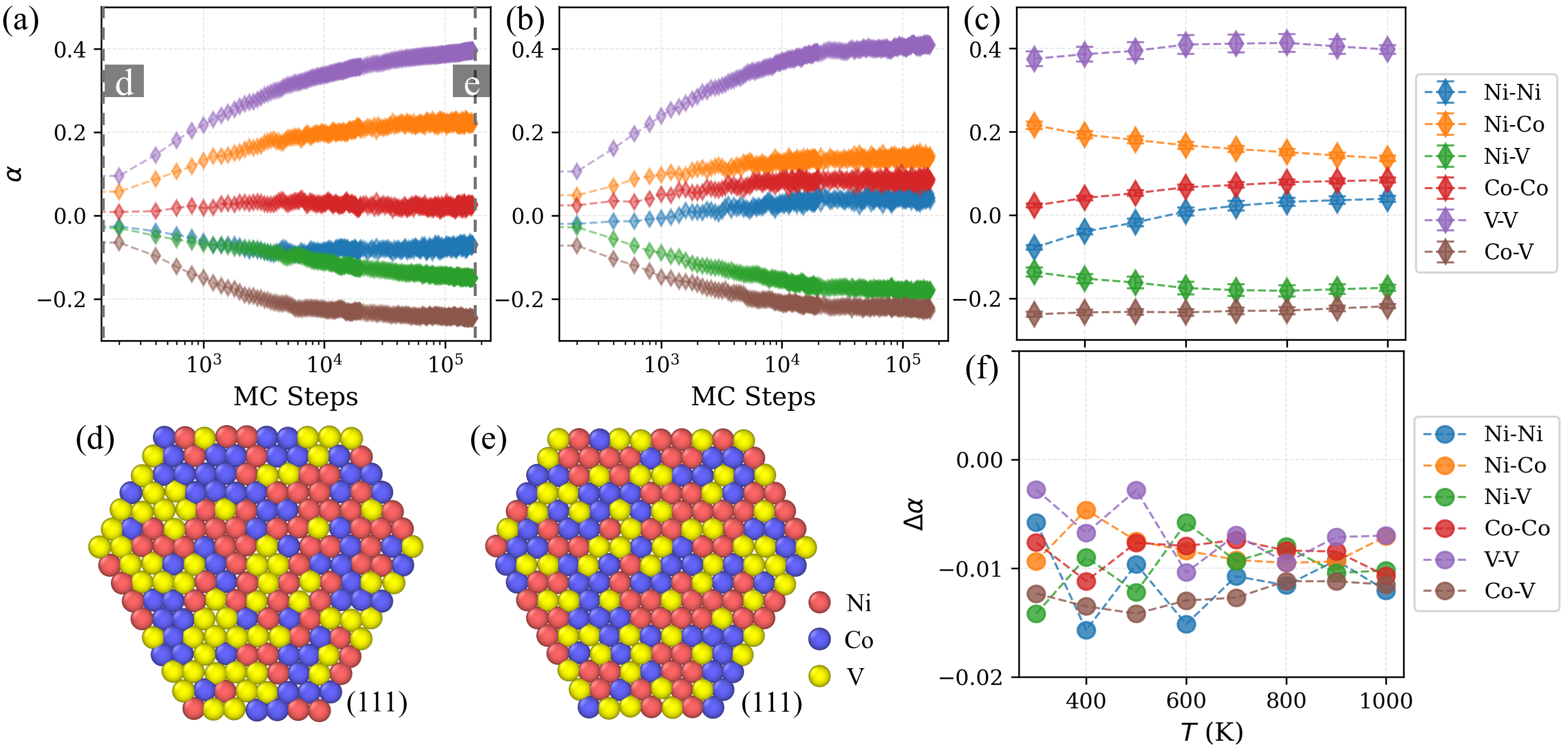}
\caption{(a–b) Evolution of first-nearest-neighbor Warren–Cowley short-range order (SRO) parameters for equiatomic CoNiV at 300 K and 1000 K as a function of MC steps, obtained from hybrid MCMD simulations. (c) Saturated Warren–Cowley parameters as a function of temperature. Error bars represent the standard deviation from six independent MC simulations. (d–e) Representative atomic configurations showing the initial random solid solution and the equilibrated CSRO state. (f) Difference in saturated Warren–Cowley parameters between systems containing 1 at.\% H and hydrogen-free systems. Negative values for Ni–V and Co–V indicate enhanced heteroatomic pairing relative to a random alloy, whereas positive values indicate avoidance of the corresponding nearest-neighbor pairs. }
\label{fig:SRO}
\end{figure}

Hybrid MC MD simulations reveal pronounced CSRO in the equiatomic CoNiV alloy from 300~K to 1000~K. Figure~\ref{fig:SRO}(a-b) summarizes the SRO evolution behavior through the Warren–Cowley parameters $\alpha_{ij}$ for first nearest-neighbors,  extracted from Monte Carlo simulations at 300~K and 1000~K. Strongly negative values of $\alpha_{\text{Ni–V}}$ and $\alpha_{\text{Co–V}}$ indicate a significantly enhanced probability of Ni–V and Co–V nearest-neighbor pairs relative to a random solid solution. In contrast, $\alpha_{\text{V-V}}$ is strongly positive (about $+0.4$), reflecting a pronounced avoidance of V–V first-nearest-neighbor bonding. 

The SRO patterns captured by the MLIP align with experimental diffuse scattering measurements and numerical predictions for NiCoV~\cite{chen2021direct}. Physically, this indicates that in CoNiV the bonding between V and the late transition metals (Ni, Co) is energetically preferred, likely driven by two factors: i) a significant size mismatch (V is much larger than Ni and Co in terms of atomic radius) and ii) electronic interaction (as an early transition metal with low $d$-band filling, V forms energetically unfavorable V–V bonds in an FCC lattice, while Ni–V and Co–V interactions allow for favorable overlap between partially filled V $d$-states and the more filled Ni/Co $d$-states). Figure~\ref{fig:SRO}(d-e) visualizes the alloy configurations for the random solution and the final CSRO state, respectively. It reveals a microstructure where V atoms are well dispersed, with much reduced V--V direct contacts, effectively forming V-centered CSRO. The strong Ni--V/Co--V ordering has important implications. CSRO can lead to local variations in lattice parameters and elastic moduli due to the different bonding strength, potentially causing nanoscale strain that impedes dislocation motion~\cite{li2019strengthening,li2025revealing}.

Figure~\ref{fig:SRO}(c) shows the temperature dependence of the saturated CSRO parameters. In the present simulations, the ordering tendencies persist across the studied temperature range, with only weak temperature dependence, especially for the V-associated pairs. This behavior contrasts with systems such as NiCoCr, where CSRO has been reported to diminish more noticeably with increasing temperature \cite{li2019strengthening}. We note that a previous study of CoNiV employing an embedded-atom method (EAM) potential reported a more pronounced reduction in the magnitude of the Warren–Cowley parameters with temperature \cite{li2025revealing}. The origin of this discrepancy is not yet fully resolved. One plausible contributing factor is the difference in interatomic potential descriptions, namely the semi-empirical EAM formulation versus the present machine-learning potential, which inherently differ in how interaction energetics are represented. At present, further studies (preferably experimental) are necessary to reconcile these observations.  

Next, we evaluate whether H presence would affect CSRO formation. Figure~\ref{fig:SRO}(f) shows the difference in saturated Warren–Cowley parameters between systems containing 1.0 at.\% H and H-free systems, defined as $\Delta \alpha = \alpha_{\text{with H}} - \alpha_{\text{no H}}$. The magnitude of $\Delta \alpha$ is on the order of 0.01 for all atomic pairs, which is negligible compared to the baseline SRO magnitude. This indicates that dilute hydrogen does not measurably alter the substitutional ordering tendencies in VCoNi. This weak coupling is not surprising: CSRO in VCoNi is primarily driven by metal–metal bonding and elastic interactions associated with size mismatch and electronic effects, while 1 at.\% hydrogen constitutes a dilute interstitial perturbation interacting locally with the neighboring atoms. Similar behavior has been reported in Fe-Ni-Cr alloys, where hydrogen up to 10 at.\% only slightly alters the intrinsic ordering preference \cite{su2025short}. This behavior contrasts with some other alloy systems, such as Pd-rich, Fe-V, and high-Mn alloys, where hydrogen has been shown to apparently promote ordering \cite{chandrasekhar2014computational,bloch2012prediction,bae2021hydrogen}. In comparison, the V-centered ordering in CoNiV is strong and less susceptible to modification by dilute hydrogen. It suggests that hydrogen-related (minor) changes in chemical ordering are unlikely to be responsible for the enhanced nanotwinning and hydrogen resistance reported in CoNiV \cite{luo2020strong}. Finally, given that hydrogen in CoNiV responds to the chemically heterogeneous environment established by metal–metal interactions, it justifies treating the bulk configuration state as fixed in subsequent hydrogen solution and trapping analyses.

\begin{figure}[!htb]
\centering
\includegraphics[width=0.5\textwidth]{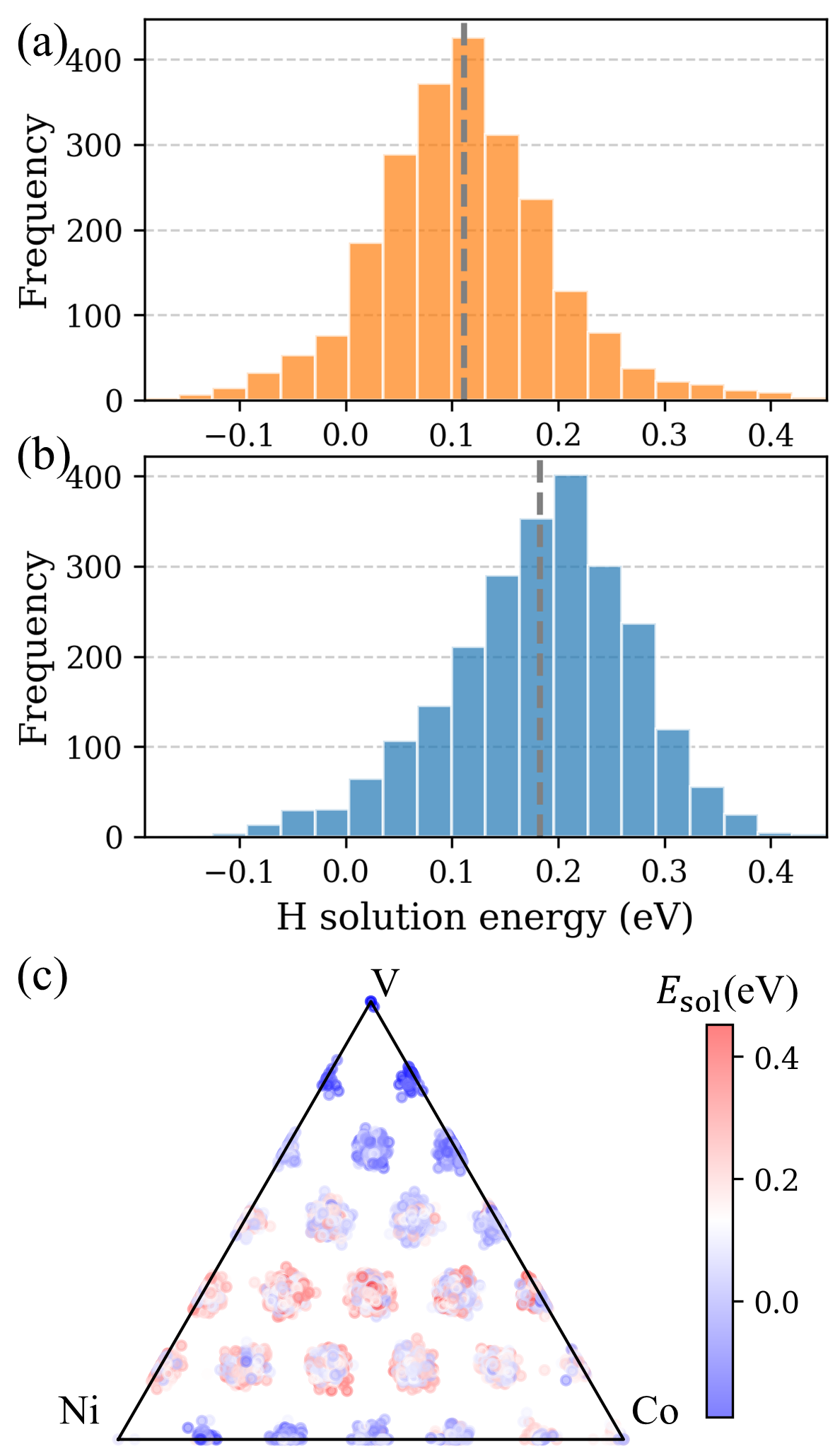}
\caption{Computed hydrogen solution energies in equiatomic CoNiV for two chemical states: (a) random solid solution and (b) CSRO configuration. (c) Ternary map showing hydrogen solution energies, where each point is positioned according to the first-nearest-neighbor composition (six surrounding metal atoms) and colored by the solution energy. A small positional jitter is applied to improve the visualization of overlapping data points.}
\label{fig:bulkH}
\end{figure}

Then, we assess the environment-dependent hydrogen solution energy, $E_{\text{sol}}$, which quantifies the thermodynamic cost of incorporating hydrogen into the lattice. As expected, hydrogen preferentially occupies the octahedral interstitial site in the FCC CoNiV structure. We computed $E_{\text{sol}}$ for two representative chemical states: a chemically random solid solution and the equilibrated CSRO configuration shown in Figure~\ref{fig:SRO}(d,e). By randomly sampling approximately 2400 octahedral sites in each configuration, the resulting distributions of $E_{\text{sol}}$ are shown in Figure~\ref{fig:bulkH}(a,b). A clear upward shift in $E_{\text{sol}}$ is observed in the CSRO state relative to the random alloy. In the random configuration, the distribution is approximately Gaussian, reflecting statistical fluctuations in local chemical environments. In contrast, the CSRO configuration exhibits a slightly left-skewed distribution. Notably, 7.8\% of sites in the random alloy exhibit negative solution energies, while this fraction decreases to 3.1\% in the CSRO configuration. The suppression of low-energy sites under CSRO contributes to an overall increase in the average solution energy, from approximately 0.11 eV in the random alloy to 0.18 eV in the CSRO alloy. Thus, the development of CSRO reduces the thermodynamic favorability of hydrogen dissolution in the lattice.

To elucidate the origin of the CSRO-induced increase in $E_{\text{sol}}$, Figure~\ref{fig:bulkH}(c) compiles all sampled octahedral sites and maps $E_{\text{sol}}$ against the local first-nearest-neighbor chemical composition (six surrounding metal atoms). It can be seen that i) the lowest solution energies strongly correlate with V-rich coordination, ii) low $E_{\text{sol}}$ sites also appear in Ni–Co environments with high Ni fraction, iii) high solution energies appear in the mixed V/Ni/Co environments with Ni+Co$\approx$(4-5)/6.  V is known to have a high affinity for hydrogen and can form different stable hydrides (e.g., VH$_x$) \cite{fukai2005metal}, so it is reasonable to expect that a site coordinating multiple V neighbors would be more energy favorable. Similar qualitative understanding applies to the Ni-Co neighboring condition with high Ni fraction due to favorable Ni-H interactions, and such a preference was found to attract H to Fe-Ni rich SRO domains in Fe-Cr-Ni alloys \cite{su2025short}. In the random alloy, statistical compositional fluctuations allow some octahedral sites to be coordinated by multiple V atoms, creating favorable incorporation sites. When CSRO develops, V–V nearest-neighbor avoidance suppresses such sites, and the preferential V-Ni and V-Co neighboring makes the local environments have a weaker affinity for hydrogen. Therefore, CSRO effectively reduces the statistical availability of deep lattice wells that would otherwise stabilize hydrogen in the lattice. From a thermodynamic standpoint, by maintaining a high degree of CSRO, through controlled thermal or mechanical processing, it may intrinsically limit bulk hydrogen uptake, which can mitigate one contributor to hydrogen embrittlement. On the other hand, an overall elevated bulk solution energy may also alter hydrogen enrichment behavior, as it would enhance the thermodynamic driving force for hydrogen to segregate to stronger traps, such as vacancies, grain boundaries, or dislocations. In this sense, CSRO may not simply reduce hydrogen uptake, but rather redistribute hydrogen toward microstructural defects where local strain fields or chemical bonding provide deeper trapping wells. For this, we next examine hydrogen interactions with dislocations in random and CSRO CoNiV.

\subsection{Hydrogen Interaction with Dislocation}
\begin{figure}[!htb]
\centering
\includegraphics[width=1.0\textwidth]{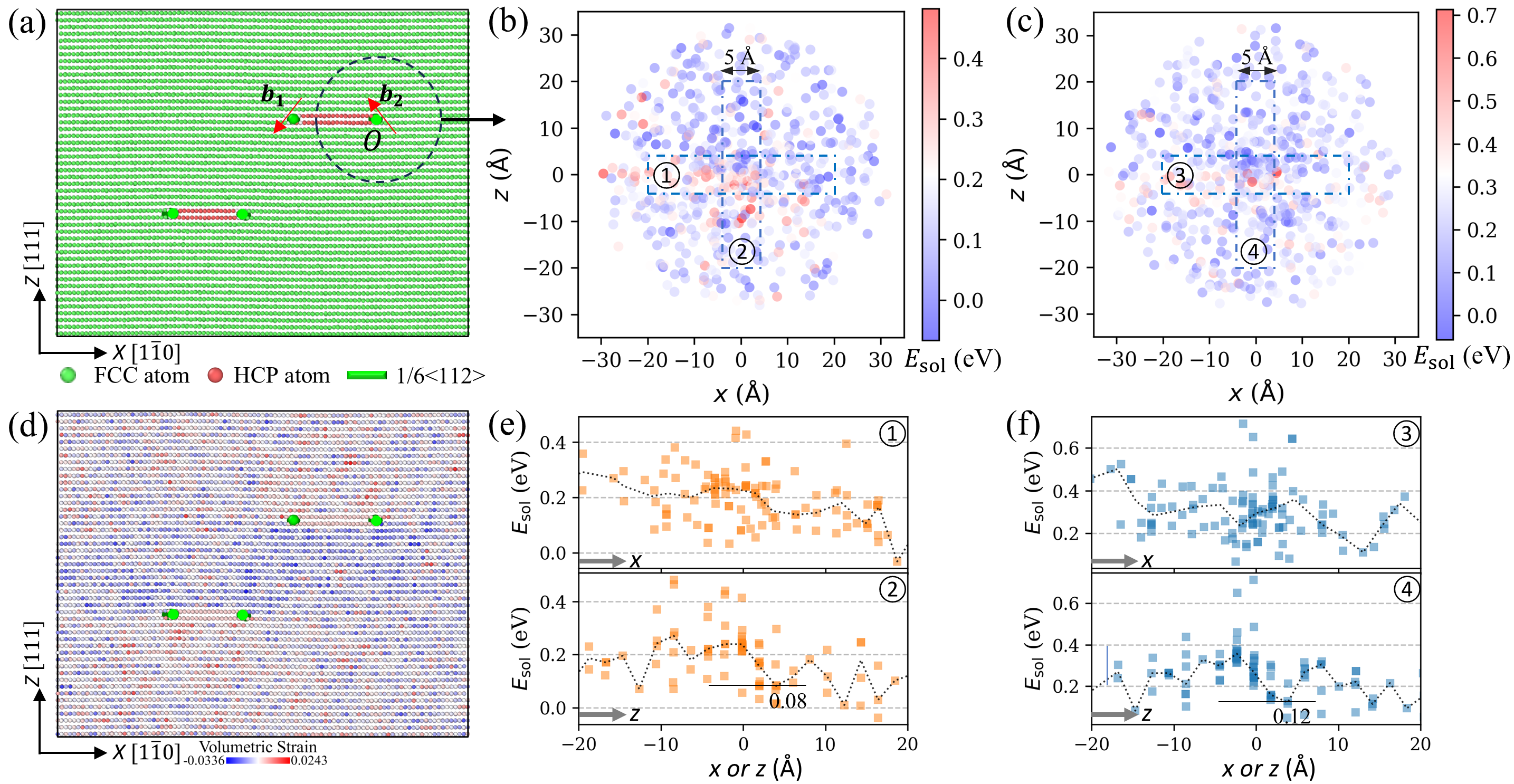}
\caption{(a) Relaxed dislocation dipole configuration in the CSRO configuration; each perfect dislocation dissociates into two Shockley partials separated by a stacking fault ribbon. (b–c) Spatial distribution of hydrogen solution energy at octahedral interstitial sites within a cylindrical region surrounding one partial dislocation line in (b) the random solid solution and (c) the CSRO configuration ($600$ data points). A small positional jitter is applied to improve the visualization of overlapping data points. (d) Hydrostatic strain field around the dislocation, showing the compressive and tensile regions associated with the stacking fault geometry. (e–f) Line profiles of hydrogen solution energy across the dislocation core along the directions indicated in (b) and (c), respectively. }
\label{fig:disloc}
\end{figure}

Dislocations in metals create traps for interstitial hydrogen \cite{chen2020observation}.To examine hydrogen–dislocation interactions, perfect dislocation dipoles were constructed in both the random solid solution and CSRO configurations. After low-temperature (10 K) dynamics relaxation of the structure, the energy-minimized configuration (Figure~\ref{fig:disloc}a) shows the $\tfrac{a}{2}\langle110\rangle{111}$ dislocation dissociated into two Shockley partials separated by a stacking fault ribbon, common in FCC metals. The corresponding hydrostatic strain map (Figure~\ref{fig:disloc}d) exhibits the characteristic compressive and tensile fields surrounding the dislocation core, which can lead to spatial variations in hydrogen segregation tendency.

Figure~\ref{fig:disloc}(b,c) maps the hydrogen solution energy at octahedral interstitial sites within a cylindrical region surrounding one partial dislocation line for random solution and CSRO configurations, respectively. In both cases, $E_{\text{sol}}$ varies strongly with position relative to the dislocation, due to the coupling between the local chemical environments and the local strain field. The tensile hydrostatic field adjacent to the dislocation core corresponds to reduced solution energies, indicating preferential segregation to this region. This tendency is more pronounced in the CSRO configuration. Quantitatively, comparing the average solution energy in the tensile core region to the bulk value yields an effective trapping strength of approximately $\sim$0.03 eV in the random alloy (0.11–0.08 eV) and $\sim$0.06 eV in the CSRO alloy (0.18–0.12 eV), based on the averaged profiles shown in Figures~\ref{fig:disloc}(e,f). Note that these values represent average trapping strengths, as the solution energy is strongly dependent on the local chemical environment. The spatial extent of the trapping region is roughly 5~\AA\ from the dislocation core, corresponding to the region of significant tensile hydrostatic strain. The calculated trapping energies indicate relatively weak, reversible trapping, much lower than vacancy trapping energies in Ni (typically 0.2–0.3 eV \cite{connetable2014segregation,nazarov2014ab}). By comparison, local chemical effects dominate over the strain contribution of the dislocation, as also reflected by the broader distribution of solution energies arising from chemical environments. In the CSRO case, at the discolcation core, the ordering is locally disrupted compared to the bulk, and the solution energies hence locally reduce close to those in the random solution. As CSRO elevates the bulk solution energy, the relative driving force for hydrogen partitioning to the dislocation core becomes more pronounced in the CSRO case. 
 
From Figure~\ref{fig:disloc}(e,f) top plots, hydrogen does not exhibit preference within the stacking fault ribbon itself. In fact, the solution energy in the faulted region is generally higher than that in the bulk lattice. This behavior can be explained. Stacking faults are primarily shear defects with minimal hydrostatic strain (Figure~\ref{fig:disloc}d), while hydrogen energetics in FCC metals are more sensitive to volumetric strain than to shear distortions. Moreover, the local atomic environment within the intrinsic stacking fault resembles HCP stacking, which does not provide stronger hydrogen stabilization compared to the FCC lattice. As a result, hydrogen trapping is dominated by core elastic fields and chemistry, not the planar defect. Given this, hydrogen is unlikely to significantly alter the stacking fault energy at dilute concentrations. Instead, its influence likely originates from interactions with dislocation cores, where it can affect core energetics and mobility, and thus plastic deformation. These effects will be examined through dynamic simulations in future work. This interpretation also informs the understanding of hydrogen resistance in CoNiV. While Luo et al. \cite{luo2020strong} proposed that hydrogen may reduce stacking fault energy and promote nanotwinning with enhanced strain hardening, the present results suggest that any hydrogen-induced changes in deformation behavior are more plausibly governed by modifications to dislocation behavior rather than direct effects on planar fault energetics.

\section{Conclusion}
In this work, we developed and validated a machine-learning interatomic potential for the Co-Ni-V-H system and used it to probe the interplay of CSRO, hydrogen energetics, and dislocation behavior. The potential well reproduces fundamental energetic and structural properties, enabling atomistic simulations beyond the scale of direct DFT. Our results confirm strong V-centered CSRO in CoNiV and demonstrate that chemical ordering significantly modifies hydrogen thermodynamics: the ordered state exhibits higher hydrogen solution energies at favorable octahedral interstitial sites compared to the random alloy. These findings suggest that CSRO intrinsically reduces bulk hydrogen uptake by suppressing strongly binding local environments. Moreover, hydrogen preferentially segregates to dislocation cores, which act as shallow, reversible traps. In comparison, chemical effects exert a stronger influence on hydrogen energetics than the dislocation strain field itself. This interplay offers an atomistic perspective on the hydrogen tolerance reported in CoNiV alloys and suggests that tuning local chemical order may provide an effective pathway to mitigate hydrogen embrittlement in CoNiV alloys. Future work will extend to dynamic hydrogen interactions with microstructural defects across varying CSRO states.

\section*{Acknowledgment} 
This work was supported by the U.S. Department of Energy, Office of Science, Office of Basic Energy Sciences, Materials Sciences and Engineering Division, under Award Number DE-SC0025170. This research used resources of the National Energy Research Scientific Computing Center, a DOE Office of Science User Facility supported by the Office of Science of the U.S. Department of Energy under Contract No. DE-AC02-05CH11231 using NERSC award BES-ERCAP0036165.

\bibliographystyle{unsrt}
\bibliography{references}
\end{document}